\documentclass[aps,twocolumn,10pt,prb,longbibliography,superscriptaddress]{revtex4-1}

\usepackage{amsmath}
\usepackage{amsfonts}
\usepackage{latexsym}
\usepackage{graphicx}
\usepackage{color}
\usepackage{bm}
\usepackage{float}
\usepackage{hyperref}
\usepackage[all]{hypcap}
\usepackage{blindtext}
\begin{document}


\title{Novel Single-mode Narrow-band Photon Source of High Brightness for Hybrid Quantum Systems} 



\author{Amir Moqanaki}
\author{Francesco Massa}
\author{Philip Walther}
\affiliation{Faculty of Physics, University of Vienna, Boltzmanngasse 5, A-1090 Vienna, Austria}


\date{\today}

\begin{abstract}

Cavity-enhanced Spontaneous parametric down-conversion (SPDC) is capable of efficient generation of single photons with suitable spectral properties for interfacing with the atoms. However, beside the remarkable progress of this technique, multi-mode longitudinal emission remains as major drawback. Here we demonstrate a bright source of single photons that overcomes this limitation by a novel mode-selection technique based on the introduction of an additional birefringent element to the cavity. This enables us to tune the double resonance condition independent of the phase matching, and thus to achieve single-mode operation without mode filters. Our source emits single-frequency-mode photons at 852 $nm$, which is compatible to the Cs D2 line, with a bandwidth of 10.9 $MHz$ and a photon-pair generation rate exceeding 47 $KHz$ at 10 $mW$ of pump power, while maintaining a low $g^{(2)}(0) = 0.13$. The efficiency of our source is further underlined by measuring a four-photon generation rate of 37 $Hz$ at 20 $mW$ of pump power. This brightness opens up a variety of new applications reaching from hybrid light-matter interactions to optical quantum information tasks based on long temporal coherence.  
\end{abstract}


\maketitle 

\section{Introduction}
\label{introduction}

Single photons' mobility, efficient detection, and ease of manipulation make them the system of choice for observing many quantum phenomena\cite{belltest, teleportation} and the natural choice for quantum information processing applications\cite{boson1, boson2, boson3, boson4, one-way}. However, the lack of photon-photon interactions raises the challenge to implement two-qubit gates\cite{klm}. One promising strategy to overcome this, is interfacing single photons with the strong optical non-linearities provided by matter-based quantum systems\cite{turchette1995measurement, schmidt1996giant, volz2014, hacker2016}. Such hybrid quantum systems, with the combined benefits of both photons and matter, can realize quantum devices such as two-qubit gates, quantum memories, quantum repeaters, and eventually a full-scale quantum network\cite{kimble2008quantum}.\\
Spontaneous parametric down-conversion (SPDC) has been widely used to generate high-purity single photons\cite{eisaman2011} at broad range of frequencies at room temperature. In contrast to other photon generation techniques\cite{senellart2017high, rodiek2017experimental}, SPDC also allows for efficient photon heralding. Recent developments in periodic poling and laser-written waveguides have triggered the realization of bright, robust and tunable photon sources\cite{tanzilli2001, fiorentino2007}. 


The bandwidth of photons that are generated via SPDC is in the order of hundreds of GHz to THz\cite{saleh-teich, rubin1994, fejer1992}, which is orders of magnitude broader than the few MHz linewidth of typical atomic transitions. One obvious method for narrowing the bandwidth is passive spectral filtering. However, this leads to significant losses, making such techniques very inefficient\cite{haase2009tunable}.\\

It has been shown that an optical parametric oscillator (OPO) pumped well below its threshold can emit single photons within the bandwidth of its cavity and greatly enhance the spectral brightness\cite{ou1999, lu2000}. Since the SPDC bandwidth is typically larger than the free-spectral range (FSR) of the enhancement cavity, the resulting source has a multi-mode spectral characteristic\cite{neergaard-nielsen2007, scholz2007, bao2008, rambach2017err}. Additional filters such as filter cavities\cite{neergaard-nielsen2007, scholz2007, bao2008, ahlrichs2013} and atomic line filters\cite{wolfgramm2011} have been employed to suppress these unwanted modes. But introducing these mode-filtering stages comes at the cost of photon loss and increased complexity of the setup.\\
Recently, doubly resonant OPOs have opened up the possibility to use the so-called \textit{cluster effect} and exploit its frequency- selectivity to enable narrow-band photon sources to perform with significantly fewer modes\cite{ahlrichs2016, foertsch2015, luo2015, rielander2013ultra}.\\
Reaching the necessary doubly resonant condition for the cluster effect is unfortunately not trivial and has been mainly attempted at highly non-degenerate SPDC\cite{foertsch2015, luo2015, rielander2013ultra}, or has not been sufficient to directly generate single mode narrow-band photons yet\cite{ahlrichs2016}.\\
In this work we report a novel approach to reach doubly resonant condition by inserting an additional birefringent crystal in a type-II OPO and tuning the clustering independent of the SPDC phase-matching. In this way, we are able to obtain a single cluster within the SPDC bandwidth and reaching single-mode operation. Tuning the cluster effect allows us to generate photons at a broad range frequencies, covering the degenerate case. The measured number of generated longitudinal modes from our source is to our knowledge the lowest ever reported. Our mode-selection scheme minimizes photon losses because of the additional filtering, and enhancing the generation rate, which even allows us to directly observe more than one photon-pair emission from our source.

\section{Single-mode operation}
\label{single-mode-operation}
\begin{figure}
	\centering
	\includegraphics[width=\columnwidth]{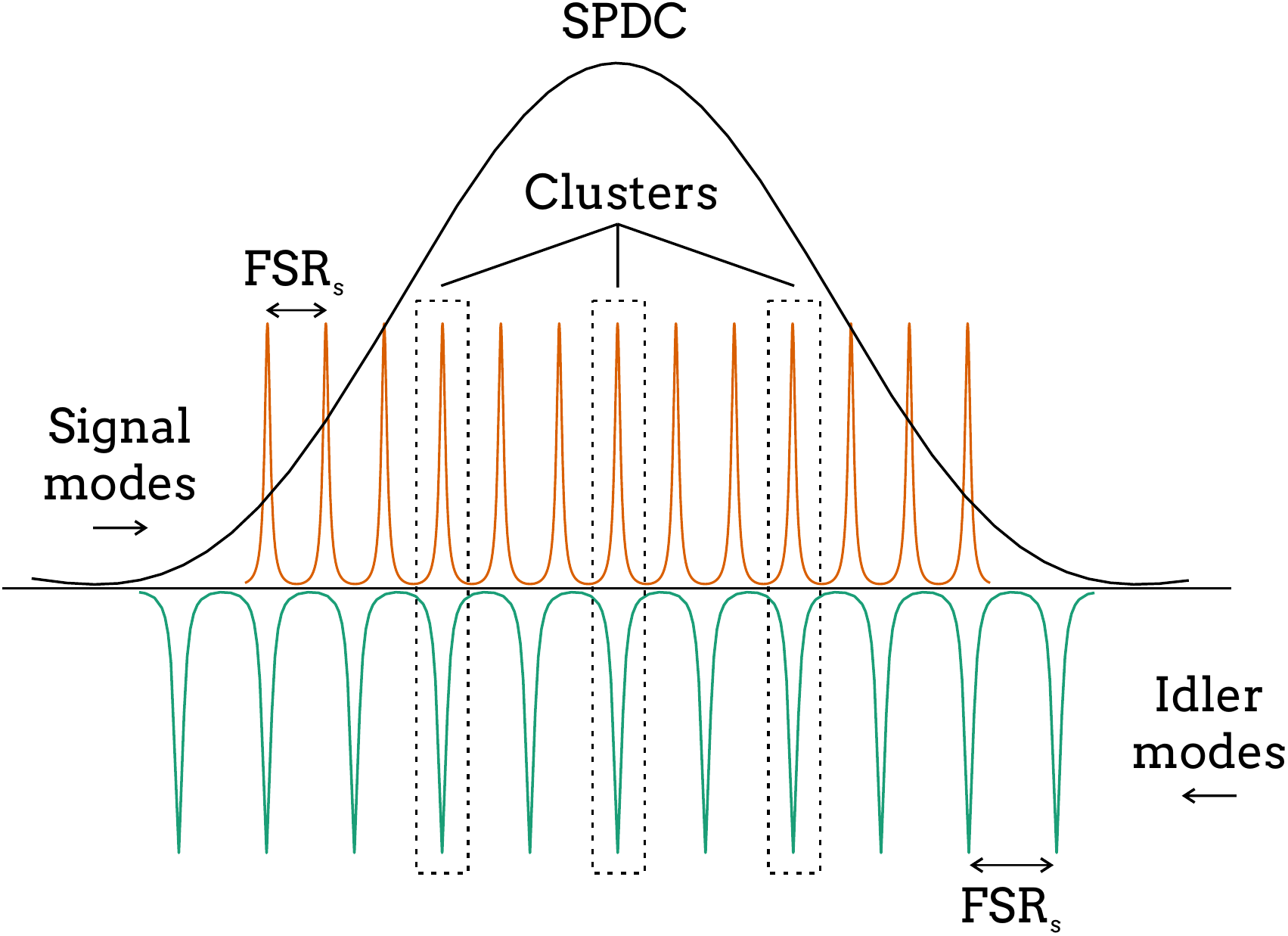}
	\caption{\footnotesize \textbf{Cluster effect.} The longitudinal modes for signal are shown in orange, and the modes for idler are in green. The two arrows indicate the direction in which frequency increases. For maximum SPDC gain, both signal and idler should be resonant at the center of the SPDC bandwidth (black solid curve), as depicted in the figure. The overlap condition is periodical, but because of the difference in the FSRs for signal and idler, the modes that are next to the fully overlapping ones coincide only partially. This leads to a spectrum that is made of clusters of modes, separated by $\Delta \nu_C$. In each cluster a bright central mode is surrounded by weaker neighboring modes. In a high finesse cavity, the linewidth of signal and idler modes are narrow enough to make the partial overlap negligible.}
	\label{fig:cluster}
\end{figure}

In a doubly resonant OPO, SPDC emission can occur only in the signal and idler longitudinal modes that are both resonant to the cavity at the same time. Due to birefringence or frequency dependence of the refractive index of optical components, signal and idler modes can have different FSRs. In this case, not every signal mode within the SPDC bandwidth is simultaneously resonant with a corresponding idler mode, and the resulting OPO spectrum is made of \textit{clusters} of few modes instead of a comb-like pattern (see Fig. \ref{fig:cluster}). This effect is known as the \textit{cluster effect}\cite{ebrahimzadeh2010cw, eckardt1991, henderson1995, nabors1990}. 

The clusters are separated by a combined FSR that is larger than both signal and idler FSRs, given by the relation below\cite{ebrahimzadeh2010cw}, in which $FSR_{s, i}$ are the FSR for the signal and the idler modes, respectively:

\begin{equation}
	\Delta \nu_c = \frac{FSR_s \cdot FSR_i}{\vert FSR_s - FSR_i \vert}.
	\label{Eq:cluster}
\end{equation}

However, reaching the doubly resonant condition at a desired wavelength is not trivial. Because tuning the cluster effect involves modifying the refractive indices of the SPDC crystal for signal and idler, consecutively affecting the SPDC phase-matching condition and complicating the process. When we add another degree of freedom dedicated for tuning the cluster separation - without affecting phase-matching - then we can simplify the tuning. Previous works have already partially investigated this by angle tuning of the SPDC crystal or applying voltage across the crystal\cite{eckardt1991, ahlrichs2016}. Yet these efforts have not been sufficient to fully suppress the unwanted longitudinal modes.\\
We consider having a linear cavity with length $d$, which contains an SPDC crystal with length $L$, a birefringent crystal - which we call \textit{tuning crystal} -, with length $L^\prime$, and some air gap ($L_{gap} = d - L - L^\prime$). We take $n_s$, $n_i$ as the group indices for signal and idler at the SPDC crystal, and $n_{s}^\prime$, $n_{i}^\prime$ as the group indices at the tuning crystal. The tuning crystal should be mounted such that the birefringence of the SPDC crystal is partially compensated, meaning that $\left(n_s - n_i\right)\left(n_s^\prime - n_i^\prime\right) < 0$. The tuning crystal length, $L^\prime$, should not be long enough to fully compensate for the birefringence of the cavity, because in this case all the signal and idler modes would completely overlap and the clustering would not occur.

By using Eq. \ref{Eq:cluster}, for the cluster separation in this arrangement we obtain:

\begin{equation}
	\Delta \nu_c = \frac{c}{2 \vert n_{s} L - n_{i} L + n_{s}^\prime L^\prime - n_{i}^\prime L^\prime \vert}.
	\label{Eq:singlemode1}
\end{equation}

The additional degrees of freedom of the tuning crystal allows us to achieve the cluster effect at the desired frequencies, independent of the phase-matching condition, and, at the same time, increase the cluster separation such that only one cluster falls within the SPDC bandwidth.\\
This happens when $\Delta \nu_c > \Delta \nu_{SPDC}$, where $\Delta \nu_{SPDC} \approx c / 2\vert n_{s} L - n_{i} L\vert$ is the half-width-half-maximum of the SPDC gain profile \cite{wolfgramm2012}. By using equation \ref{Eq:singlemode1}, this condition becomes:

\begin{equation}
	\frac{1}{2} \frac{\vert n_s - n_i \vert}{\vert n^\prime_{s} - n^\prime_{i} \vert} \leq \frac{L^\prime}{L} < \frac{\vert n_s - n_i \vert}{\vert n^\prime_{s} - n^\prime_{i} \vert}.
	\label{Eq:singlemode2}
\end{equation}

This leads to emission of a single cluster which can still contain more than one mode (See Fig. \ref{fig:cluster}). When the bandwidth of the modes is narrow enough such that the partial overlap determining the neighboring modes is negligible, each cluster contains only one mode. If we define $\Delta \nu_s$ and $\Delta \nu_i$ the bandwidth of signal and idler modes, respectively, which we can write this condition as: 
\begin{equation}
	\frac{1}{2} \Delta \nu_{s} + \frac{1}{2} \Delta \nu_{i} < \vert FSR_i - FSR_s \vert.
	\label{Eq:min_separation}
\end{equation}

Approximating the same finesse, $F$, for both signal and idler, we set a minimum threshold for the finesse:

\begin{equation}
	F > \frac{1}{2} \frac{FSR_i + FSR_s}{ \vert FSR_i - FSR_s \vert}.
	\label{Eq:singlemode3}
\end{equation}

\section{Experimental realization}
\begin{figure}
	\centering
	\includegraphics[width=\columnwidth]{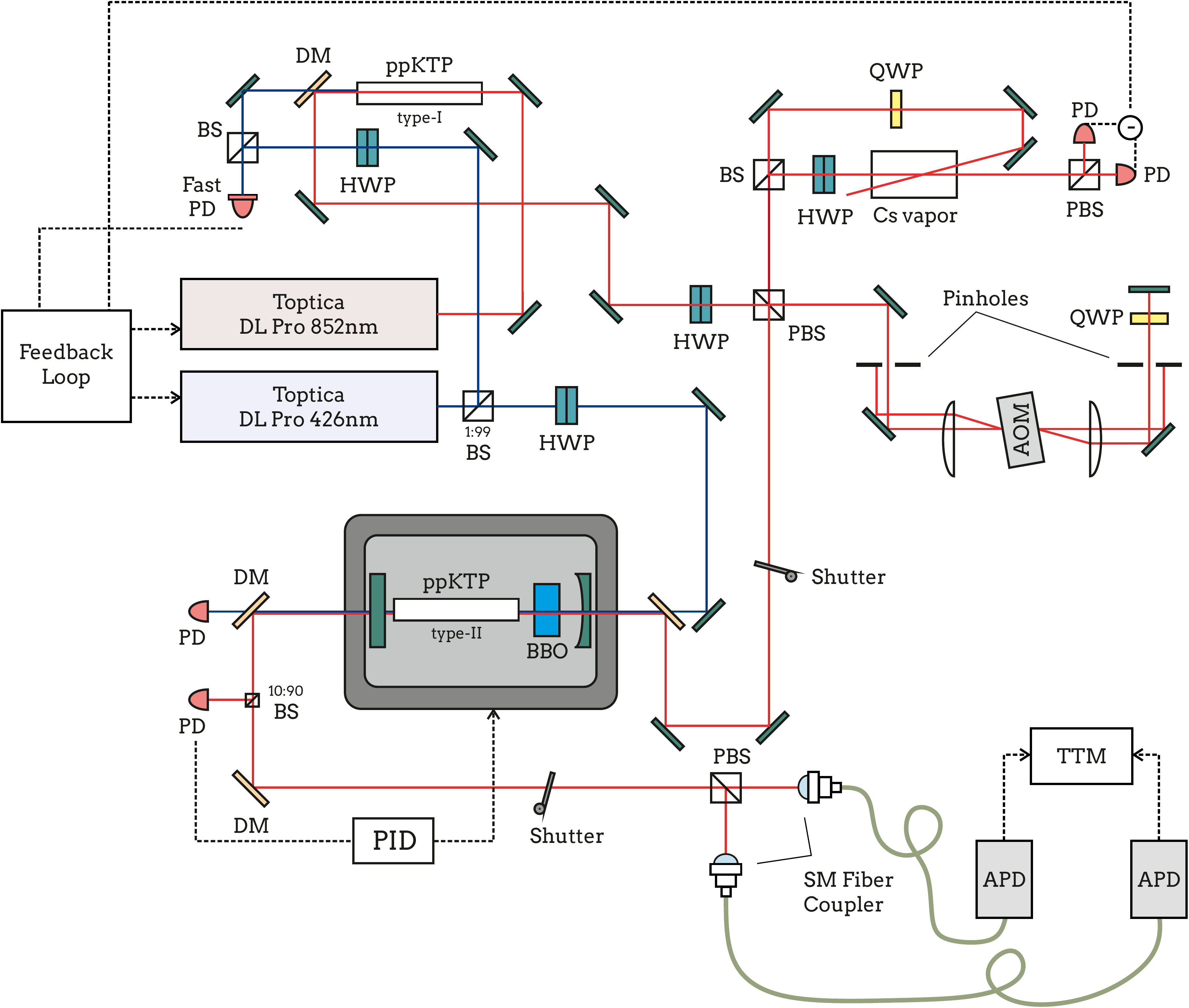}
	\caption{\footnotesize \textbf{Narrow-band single photon source.} Two tunable diode lasers operating at 852 $nm$ -called \textit{probe}, stabilized to Cs D2 line via polarization spectroscopy- and 426 $nm$ -called \textit{pump}, locked to the probe by frequency-offset lock- constitute the laser system. The OPO cavity's geometry is semi-hemispherical, and it contains two crystals: A 30-$mm$-long PPKTP (periodically-poled potassium titanyl phosphate) for collinear, type-II phase-machted SPDC of 426 $nm$ to 852 $nm$ and a 15-$mm$-long BBO (barium borate) for tuning the cluster effect. Two optical shutters switch the probe beam and block the APDs to iterate between the cavity lock and the measurement mode with 60\% duty cycle. The photons are coupled into single-mode fibers and routed to the APDs and the TTM for analyzing the temporal correlations.
	\label{fig:setup}}
\end{figure}

We show a schematic of the experimental realization of the source in Fig.\ref{fig:setup}. The probe laser (Toptica DL Pro 852, 50 $mW$) is tuned to the Cs D2 line and acts as the frequency reference. In order to lock the pump laser (Toptica DL Pro 426, 30 $mW$) to the probe laser, we frequency-double the probe by a single-pass type-I PPKTP crystal (30mm, Raicol Crystals) to 426 $nm$. We sample part of the pump laser beam and combine it with the frequency-doubled probe laser on a 50:50 BS. The resulting beat note is then detected by a fast photo-diode (Alphalas UPD-200-UP) and sent to the home-built feed-back electronics to generate the error signal and lock the pump laser to the probe laser via frequency-offset lock\cite{schunemann1999simple}. We recycle the probe laser after passing the frequency-doubling stage, and send small part of it to a double-pass AOM (Gooch \& Housego 70 MHz) arrangement to provide detuning from the desired atomic transitions. The frequency-shifted beam is used to perform polarization spectroscopy on Cs vapor to lock the probe laser to the Cs D2 line transitions. With this technique, we can stabilize both pump and probe laser frequencies to the Cs D2 line and fine tune their frequencies independently. The rest of the probe is sent to the cavity for locking.\\
We have built our OPO in a linear, semi-hemispherical configuration. The asymmetric geometry allows for non-degeneracy in transverse modes, which simplifies the alignment and coupling to single-mode fibers. The out-coupling mirror is partially reflective for the pump beam (specified R $<$ 0.97 at 426 $nm$) and has a modest transmission for the probe (T $\approx$ 0.03 at 852 $nm$). The in-coupler has a high reflection for the probe (R $>$ 0.999 at 852 $nm$) and partial reflection for the pump (specified R $<$ 0.57 at 426 $nm$).\\
We use a 30-$mm$-long, type-II PPKTP crystal (Raicol Crystals) for SPDC and a 15-$mm$-long BBO (Caston Inc) for tuning the cluster separation. Both crystals are anti-reflection coated for 426 $nm$ and 852 $nm$. We control the crystals temperature with a precision better than 2 $mK$ using home-built crystal ovens and electronics from Team Wavelength (PTC-5K-CH). The whole crystal assembly is mounted on a 5-axis stage (Thorlabs PY005) for fine alignment of the OPO cavity.\\
The cavity is about 60 $mm$ long from which we estimate $FSR_s \approx$ 1.56 $GHz$ and $FSR_i \approx$ 1.58 $GHz$. According to Eq. \ref{Eq:singlemode3}, the minimum finesse to achieve single-mode operation is 81. Considering the mentioned mirrors reflectivities and lumped transmission loss of 0.01 per crystal, we estimate an average finesse of 121, which is safely above this limit. We separate the single photons from  the residual pump beam by using a combination of two dichroic mirrors, two long-pass filters (Thorlabs FEL0750), and one band-pass filter (Thorlabs FBH850-40). We make sure the cavity is resonant for both signal and idler by monitoring the overlap of the transmission peaks of the probe for the orthogonal polarization modes.\\
A 10:90 BS samples part of the probe for locking the cavity. The transmitted probe beam is sent to a digital lock module (Toptica Digilock) for a side-of-the-fringe lock. We use two optical shutters (Thorlabs SHB05T) controlled by an Arduino Uno micro-controller to block the probe beam during the measurement cycles (600 $ms$) and to prevent it from reaching the detectors. After separating signal and idler by a PBS, we couple the photons into single-mode fibers (53\% average coupling efficiency), and send them to the single-photon counting modules (Excelitas SPCM-ARQH-12-FC) for detection and analysis. Our detectors have a typical FWHM jitter of 350 $ps$ and a detection efficiency of about 69\%. The detectors read-out is recorded by a time-tagging module (AIT TTM8000).\\
\label{experimental}

\section{Characterization and Results}
\begin{figure}
	\centering
	\includegraphics[width=\columnwidth]{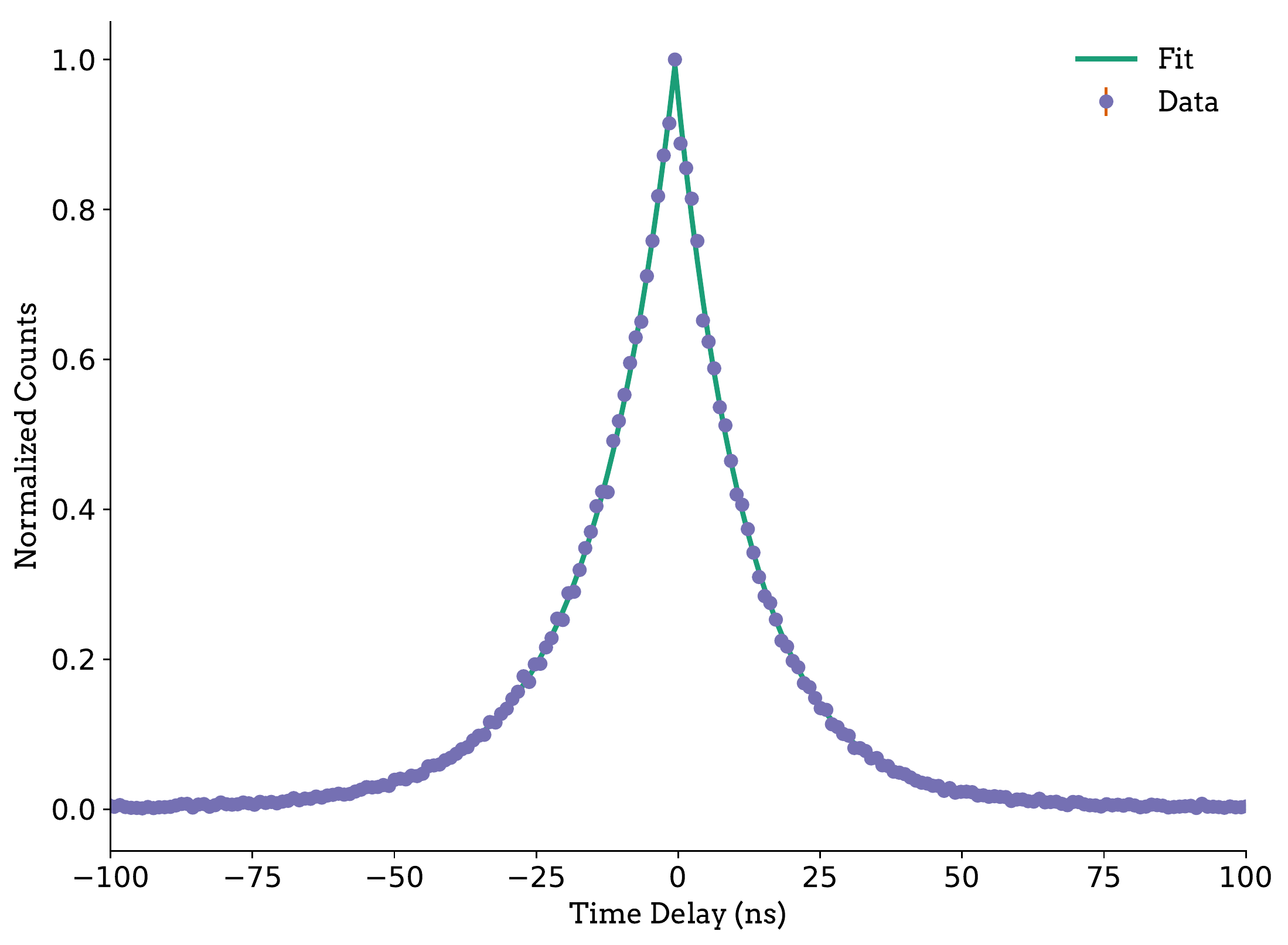}
	\caption{\footnotesize \textbf{Experimental cross-correlation.} The delays are calculated from the coincidence time-tags recorded by the TTM. We normalize the counts to the maximum and correct for the accidentals and the dark counts.  Each time bin has a width of 2.97 $ns$, which corresponds to six times the combined jitter of the detectors. The errorbars are calculated from the Poisson statistics of the photon counting. The fit function is an exponential (See Eq. \ref{Eq:xc}) and the decay rates $\gamma_s$ and $\gamma_i$ are the fit parameters. The FWHM of the fit is (18.7 $\pm$ 0.5) $ns$, resulting into a bandwidth of (10.9 $\pm$ 0.3) $MHz$.
	\label{fig:xc}}
\end{figure}

We study the temporal correlation functions of the emitted photons in order to infer their spectral properties and to demonstrate the single-photon character of our source\cite{scholz2009statistics, clausen2014, ahlrichs2016}. The cross-correlation function $G^{(1,1)}(\tau)$ is proportional to the probability of detecting a signal photon at time $t$ and an idler photon at time $t + \tau$. The FWHM of the cross-correlation temporal profile is defined as the cross-correlation time, $\tau_c$, and is inversely proportional to the bandwidth of the emitted photons. For a doubly resonant type-II OPO, we have\cite{scholz2009statistics}:

\begin{equation}
	G^{(1,1)}(\tau) \approx H(\tau) e^{-2 \gamma_s \tau} + H(-\tau ) e^{2 \gamma_i \tau},
\label{Eq:xc}
\end{equation}

where $H(\tau)$ is the Heaveside step function and $\gamma_{s,i}$ are the signal and idler decay rates in the cavity. We plot the measured cross-correlation in Fig. \ref{fig:xc}, together with a fit from Eq. \ref{Eq:xc}. This cross-correlation signal is not symmetric, because of the birefringence in the OPO, which is not fully compensated by the tuning crystal. The cross-correlation time, calculated from the fit, is (18.7 $\pm$ 0.5) $ns$. Considering the doubly-resonant condition, the effective bandwidth of the photons is given by $\frac{0.64}{\pi\tau_c}$, which corresponds to a bandwidth of (10.9 $\pm$ 0.3) $MHz$ for the photon-pairs\cite{scholz2009analytical, scholz2009statistics}.\\

\begin{figure}
	\centering
	\includegraphics[width=\columnwidth]{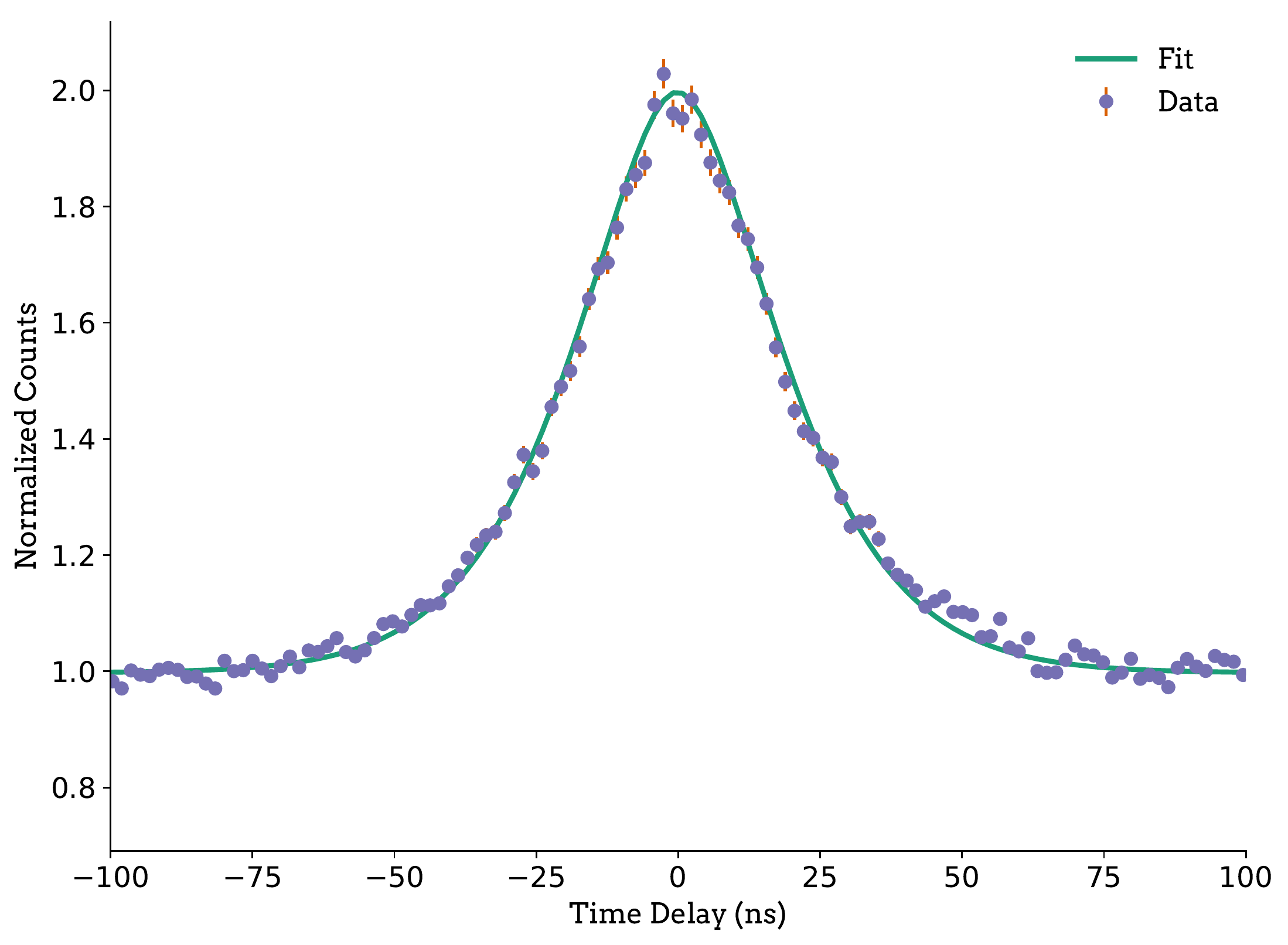}
	\caption{\footnotesize \textbf{Experimental auto-correlation.} The delays are calculated from the coincidence time-tags recorded by the TTM. We normalize the counts to the value at far delays. The time bin size is 4.95 $ns$, which corresponds to ten times the combined jitter of our detectors. The errorbars are calculated from the Poisson statistics of the photon counting. The fit function is that of Eq. \ref{Eq:ac} and the fit parameters are $\gamma_ s$ and $\gamma_i$. The auto-correlation is symmetric and visibly broader than the cross-correlation.
	The auto-correlation time is calculated from the fit to be 41.1 $\pm$ 0.5 $ns$, which is more than two times larger than the cross-correlation time, as expected.
	 The peak value at 0 delay is 1.995 $\pm$ 0.001, corresponding to single-frequency-mode operation.
	\label{fig:ac}}
\end{figure}

The auto-correlation function, $G^{(2)}(\tau)$ provides information about the statistics of the single SPDC fields. We measure this by splitting the signal arm with a 50:50 fiber beam splitter (Thorlabs TW850R5F1) and by recording the coincidences at the output ports at the time delay $\tau$, thus realizing an Hanbury Brown-Twiss interferometer. Our results are plotted in Fig. \ref{fig:ac}.\\
Unlike the cross-correlation, $G^{(2)}(\tau)$ is symmetric and has a more complex shape (see Eq. \ref{Eq:ac}). Ideally, the auto-correlation would peak at $G^{(2)}(0) = 2$ when being measured by a detection system with no time-jitter, thus revealing the thermal statistics of the SPDC fields. However, given the relatively large combined jitter in the detectors (0.495 $ns$) with respect to the correlation time (18.7 $ns$), this holds true only for a single-frequency-mode photon source, and in the multi-mode case the peak value decreases, $G^{(2)}(0) < 2$. In fact, the value of auto-correlation function at 0 delay gives the effective number of modes that are present in the cavity, $N$\cite{clausen2014}:

\begin{equation}
	G^{(2)}(0) = 1 + \frac{1}{N}
	\label{Eq:modes}
\end{equation}

The general expression for the auto-correlation when the signal and the idler experience different decay rates, $\gamma_s \neq \gamma_i$, is given by\cite{luo2015}:

\begin{equation}
	G^{(2)}(\tau) \approx 1 + \Bigg[ e^{-\frac{\gamma_s + \gamma_i}{2} \lvert \tau \rvert} \bigg( 1 + \frac{\gamma_i + \gamma_s}{2} \lvert \tau \rvert \bigg) \Bigg]^2.
	\label{Eq:ac}
\end{equation}

By fitting the Eq. \ref{Eq:ac} to our experimental data, we obtain $G^{(2)}(0) =$ 1.995 $\pm$ 0.001. The errors are based on Poisson statistics and are obtained by Monte Carlo simulation of several runs\cite{montecarlo}. The auto-correlation function is expected to be broader than the cross-correlation function. We calculate the FWHM of the fit to be 41.1 $\pm$ 0.5 $ns$, more than two times larger than $\tau_c$ as expected from the theoretical predictions. These results indicate that our source operates in a single-longitudinal mode\cite{clausen2014, luo2015}.

\begin{figure}
	\centering
	\includegraphics[width=\columnwidth]{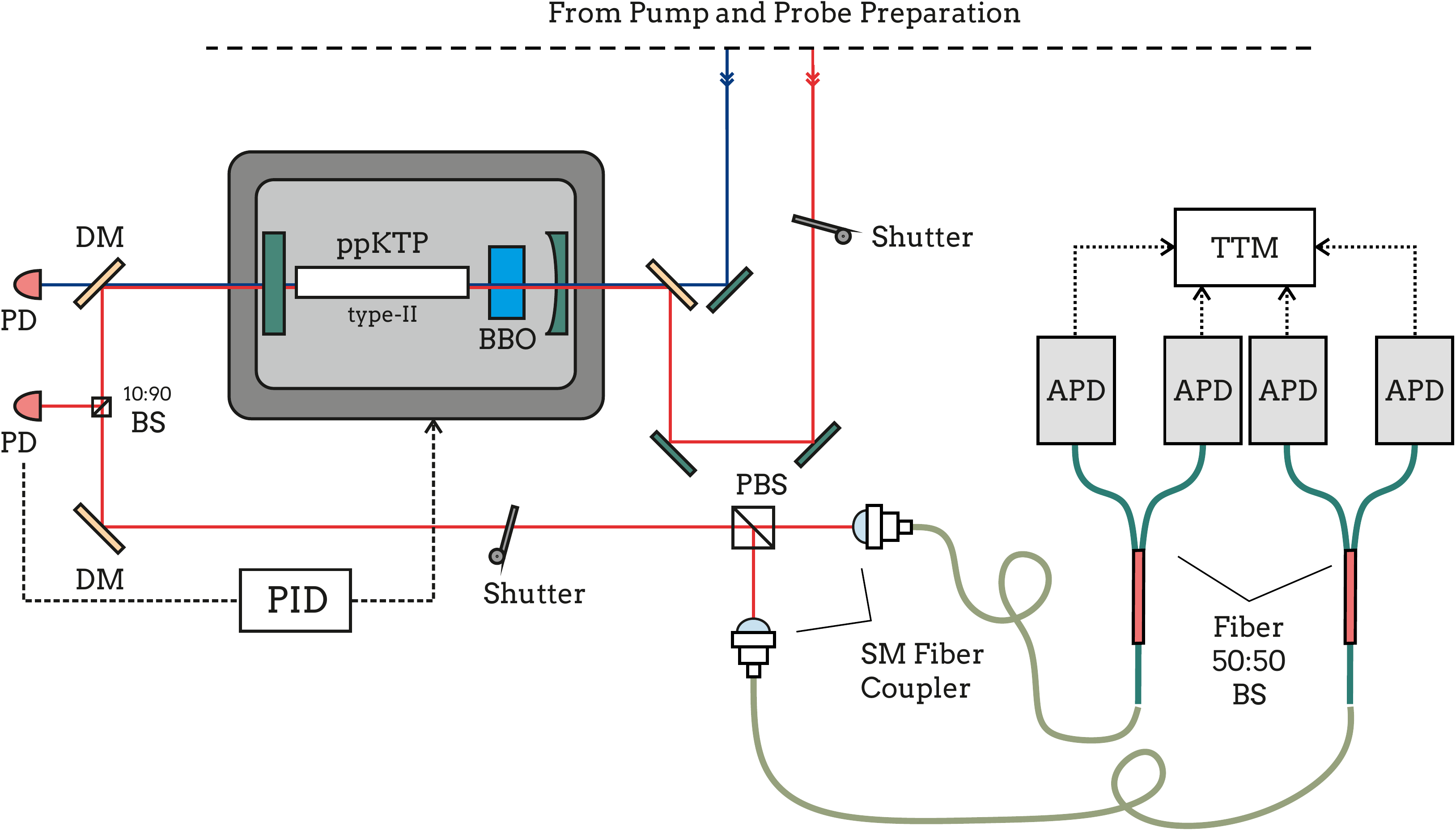}
	\caption{\footnotesize \textbf{Set-up for the measurement of the 4-fold coincidences.} For the 4-fold coincidence measurement the pump power is increased and the double pairs are spatially separated by two 50:50 fiber beam-splitters. The 4-fold coincidences are post-selected by the TTM.
	\label{fig:setup-multi}}
\end{figure}

Another parameter that is used for source characterization is the heralded or conditional second-order correlation function $g^{(2)}(\tau)$, which is measured in the same way as the auto-correlation, with the only difference that the detection of the signal photon is heralded by the corresponding detection of an idler photon. The value of this function at zero delay, $g^{(2)}(0)$ quantifies multi-photon emission from an SPDC source\cite{wolfgramm2012, rambach2017err}. This value should be $g^{(2)}(0) = 0$ for an ideal single-photon source without higher order emissions. We calculate $g^{(2)}(0)$ as\cite{wolfgramm2012}:

\begin{equation}
	g^{(2)}(0) = \frac{2 \times \text{C}_{\text{H}} \times \text{CC}_{\text{HAB}}}{(\text{CC}_{\text{HA}} + \text{CC}_{\text{HB}})^2},
\end{equation}

where $\text{C}_{\text{H}}$ is the rate of single counts for the heralding photons, $\text{CC}_{\text{HA}}$, $\text{CC}_{\text{HB}}$ are the 2-fold coincidence rates between the heralding photon and the two arms of the the Hanbury Brown-Twiss interferometer respectively, and $\text{CC}_{\text{HAB}}$ is the 3-fold coincidence rate. We measure the heralded auto-correlation at 1 $mW$ of pump power to be $g^{(2)}(0) =$ 0.04 $\pm$ 0.01, and at 10 $mW$ of pump power to be $g^{(2)}(0) =$ 0.130 $\pm$ 0.006. The errors are calculated from the Poisson statistics on photon counting.\\
We record 2.5 $KHz$ 2-fold coincidence counts at 10 $mW$ of pump power. This corresponds to a photon-pair generation rate of 47.5 $KHz$ after correcting for the coupling and detection loss, and the transmission of the dichroic filters, and thus to a spectral brightness of 436 $s^{-1} mW^{-1} MHz^{-1})$. This single-mode photon-pair generation rate surpasses previous results based on bulk optics and additional mode filters at the same level of photon purity\cite{rambach2017err, ahlrichs2016, tsai2018, niizeki2018}.\\
Our mode selection technique allows for direct single-mode emission, and therefore does not require any additional filtering stage. This drastically reduces the losses and gives us the possibility to operate the source at a new parameter regime such that multi-photon generation can be detected. In order to demonstrate this, we increase the pump power to 20 $mW$ and introduce an additional 50:50 BS (see Fig. \ref{fig:setup-multi}) for being able to measure the 4-fold coincidences by probabilistically separating the photons into four output modes. At this pump power we measure a 4-fold coincidence rate of 0.28 $Hz$, and obtain a double-pair generation rate of 37 $Hz$ after correcting for losses.
\label{results}

\section{Conclusions}
Here we experimentally demonstrate a novel technique to efficiently generate narrow-bandwidth single-frequency-mode photons. We characterize our source by analyzing the temporal correlations of the emitted photons. Furthermore, we show that our mode-selection approach reduces the losses to the level that makes multi-photon generation accessible to experimental investigation. Our work opens up the possiblity to use narrow-band multi-photon states for applications such as verification of matter-based two-photon gates\cite{volz2014, hacker2016}, multi-boson correlation sampling and richer temporal interference landscapes\cite{tamma2015multiboson}, and generation of highly entangled narrow-band states for quantum communication and quantum information processing\cite{yang2009experimental}.
\label{conclusions}

\begin{acknowledgments}
We would like to thank Kai-Hong Luo, Andreas Ahlrichs, Oliver Benson, and Morgan Mitchell for the useful discussions. We thank Fabian Laudenbach and the Austrian Institute of Technology (AIT) GmbH for kindly lending us their TTMs. We would like to thank European Commission under the project ErBeStA (No. 800942), Austrian Research Promotion Agency (FFG) through the QuantERA ERA-NET Cofund project HiPhoP, Austrian Science Fund (FWF) through the doctoral programme CoQuS (W1210), NaMuG (P30067-N36), and GRIPS (P30817-N36), University of Vienna via the Research Platform TURIS, and Red Bull GmbH for their generous support.
\end{acknowledgments}


%
%

%

\bibliography{narrowband_refs}

\end{document}